*Article*

# Effect of the Protein Corona Formation on Antibody Functionalized Liquid Lipid Nanocarriers

Saul A. Navarro-Marchal [1,2,3], Marina Martín-Contreras [4], David Castro-Santiago [4], Teresa del Castillo-Santaella [5,6], Pablo Graván [1,2,3,6,7], Ana Belén Jódar-Reyes [3,4,6], Juan Antonio Marchal [1,2,3,7] and José Manuel Peula-García [6,8,*]

[1] Biopathology and Regenerative Medicine Institute (IBIMER), Centre for Biomedical Research (CIBM), University of Granada, 18100 Granada, Spain; navarrosa@ugr.es (S.N.M.); gravan@ugr.es (P.G.); jmarchal@ugr.es (J.A.M.)
[2] Instituto de Investigación Biosanitaria de Granada (ibs.GRANADA), 18012 Granada, Spain
[3] Excellence Research Unit Modeling Nature (MNat), University of Granada, 18071 Granada, Spain; ajodar@ugr.es (A.B.J.R.)
[4] Department of Applied Physics, Faculty of Sciences, University of Granada, 18071 Granada, Spain; marinamctr@gmail.com (M.M.C.); david.casant@gmail.com (D.C.S.)
[5] Department of Physical Chemistry, Faculty of Pharmacy, University of Granada, 18011 Granada, Spain; tdelcastillo@ugr.es (T.C.S.)
[6] Biocolloid and Fluid Physics Group, Faculty of Sciences, University of Granada, 18071 Granada, Spain
[7] Department of Human Anatomy and Embryology, Faculty of Medicine, University of Granada, 18016 Granada, Spain
[8] Department of Applied Physics II, University of Malaga, 29071 Malaga, Spain
* Correspondence: jmpeula@uma.es (J.M.P.G.)

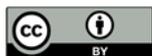

**Abstract:** The main aim of this study is to report basic knowledge on how a protein corona (PC) could affect or modify the way in which multifunctionalized nanoparticles interact with cells. With this purpose, we have firstly optimized the development of a target-specific nanocarrier by coupling a specific fluorescent antibody on the surface of functionalized lipid liquid nanocapsules (LLNCs). Thus, an anti-HER2-FITC antibody ($\alpha$HER2) has been used, HER2 being a surface receptor that is overexpressed in several tumor cells. Subsequently, the in vitro formation of a PC has been developed using fetal bovine serum supplemented with human fibrinogen. Dynamic Light Scattering (DLS), Nanoparticle Tracking Analysis (NTA), Laser Doppler Electrophoresis (LDE), and Gel Chromatography techniques have been used to assure a complete physico-chemical characterization of the nano-complexes with (LLNCs-$\alpha$HER2-PC) and without (LLNCs-$\alpha$HER2) the surrounding PC. In addition, cellular assays were performed to study the cellular uptake and the specific cellular-nanocarrier interactions using the SKBR3 (high expression of HER2) breast cancer cell line and human dermal fibroblasts (HDFa) (healthy cell line without expression of HER2 receptors as control), showing that the SKBR3 cell line had a higher transport rate (50-fold) than HDFa at 60 min with LLNCs-$\alpha$HER2. Moreover, the SKBR3 cell line incubated with LLNCs-$\alpha$HER2-PC suffered a significant reduction (40%) in the uptake. These results suggest that the formation of a PC onto LLNCs does not prevent specific cell targeting, although it does have an important influence on cell uptake.

**Keywords:** active targeting; breast cancer; cellular uptake; lipid liquid nanocapsules; protein corona

## 1. Introduction

Nanomedicine is defined as the design and development of therapeutic and/or diagnostic agents with nanoscale dimensions (with sizes ranging from 1 to 1000 nm) [1]. Many different nanoparticles (NPs) have been designed and approved for clinical use in the last few decades. Anti-cancer drug nanocarriers are one promising example of these NPs. The development of innovative therapies, such as the use of nanomedicines, is presented as an alternative to conventional chemotherapy to achieve greater safety and effectiveness in the treatment of cancer [2]. Two different strategies can be followed in order to target tumor cells: passive targeting and active targeting. Passive targeting is based on the enhanced permeability and retention effect (EPR) of tumors. The EPR phenomenon involves the accumulation of NPs at the therapeutic target without specific recognition of tumor receptors. This is due to tumors presenting abnormal vasculature, with interstices between endothelial cells averaging between 10 and 500 nm in size, accompanied by defective drainage by the lymphatic system. As a result, NPs accumulate preferentially in tumors rather than in healthy tissues [3].

Active targeting also relies on the EPR effect, but, in addition, it takes advantage of the ability of the NP's surface to bind molecules that recognize over-expressed markers on tumor cells in a selective manner. The binding of ligands that recognize specific receptors to the surface of nanoparticles is known as functionalization [3]. There are numerous recognition molecules with which NPs can be functionalized, antibodies (or fragments of antibodies) being an interesting option because they recognize any antigen in a highly specific way. For instance, the monoclonal antibody Trastuzumab (Herceptin®, Roche Pharma, Grenzach-Wyhlen, Alemania) recognizes the HER2/neu receptor (human epidermal growth factor receptor 2), which is overexpressed in multiple cancers, such as breast, lung and ovarian cancer [4]. In the case of breast cancer (BC), it is estimated that between 15 and 20% of tumors show overexpression of this receptor [5] and this overexpression is associated with worse disease progression and a higher likelihood of relapse [6]. Active targeting, achieved through vectorized nanosystems, enhances the drug internalization into tumor cells by facilitating the entry through receptor-mediated endocytosis [7]. Selective targeting of tumor cells allows for increased cytotoxicity in these cells, as well as reduced drug internalization in healthy tissues, thereby reducing side effects.

Regardless of the procedure of targeting, when the NP reaches a biological fluid, the rapid adsorption of biomolecules on its surface occurs, forming a kind of "corona". As these biomolecules are mostly proteins, this generated shell is termed as "protein corona" (PC) [8]. A PC is highly dynamic and can be divided into two entities: the "hard corona" constituted by proteins that exhibit a high affinity for the NP and establish strong interactions with it; and the "soft corona" made up of proteins that show a low affinity for the NP and whose binding is reversible [9].

This PC gives the NPs a new biological identity, and it can affect the nanoparticles' stability, biodistribution, toxicity, cellular uptake, and interaction with the immune system, among other factors [2,8,9]. Consequently, when designing a therapeutic system based on NPs, it is essential to study how it may be affected as a result of the PC formation. The material characteristics (size, surface roughness, charge, and chemistry) and environmental parameters (composition of the biological medium and chemico-physical conditions such as temperature, pH, electrolytes, and time) influence the PC formation and how this protein structure organizes around NPs [10].

In the case of active targeting, it should be checked whether the formation of a PC involves the masking of the specific ligands on the NP surface that alters its selectivity or competes in the interaction with cell receptors [11,12].

Despite these drawbacks, the design of artificial PCs with controlled physico-chemical properties has been recently described in the literature. Controlled and stable coronas offer the unexpected possibility to preserve stealth properties of designed nanocarriers, regulating its cellular interactions in physiological media and minimizing the possible consequences derived from a natural PC such as alteration of targeting efficiency, short life time by mononuclear phagocyte system (MPS) sequestration, and colloidal aggregation (Figure 1) [13–17].

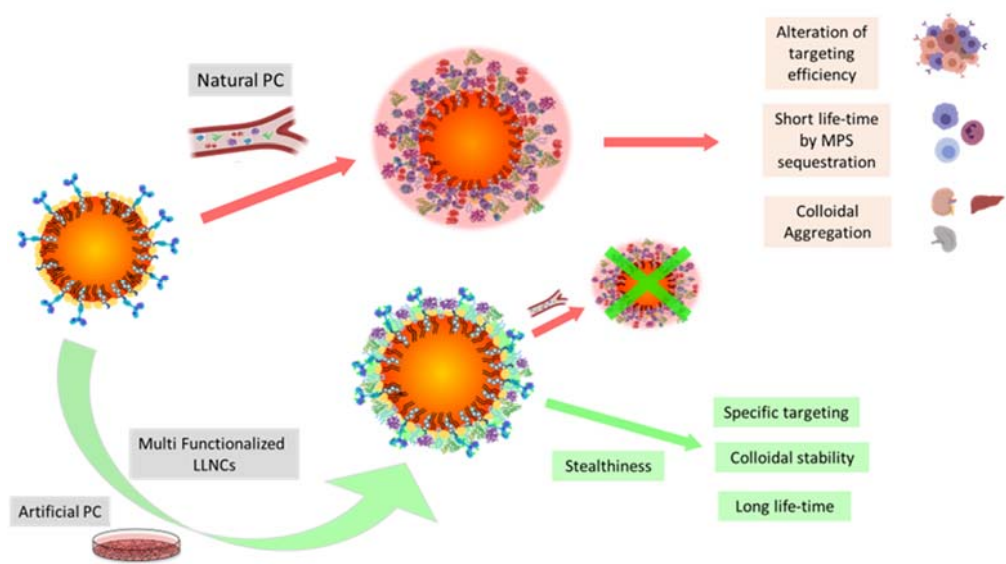

**Figure 1.** Schematic representation with different possibilities of protein corona conformed on multifunctionalized LLNCs. Red arrows indicate a negative effect and green arrows and cross indicate a positive effect.

The aim of this work is to evaluate how PC formation influences the selective recognition of the HER2 receptor by NPs consisting of olive oil. These NPs belong to the group called lipid liquid nanocapsules (LLNCs), because their general structure consists of a liquid lipid core at room temperature coated by a polymeric shell, which provides stability to the system [18]. LLNCs show great potential as therapeutic agents against cancer, as they are able to efficiently encapsulate chemotherapeutic drugs (which are usually lipophilic, and thus have very limited solubility in physiological fluids) and protect them from degradation by external agents (light, pH, and the presence of enzymes) [19]. Olive oil is also a suitable choice to form the core of LLNCs, not only because it is biocompatible, biodegradable, and non-toxic [20] but also because it exhibits some intrinsic antitumor activity [21].

We successfully developed monodisperse LLNCs with diameters ranging from 100 to 200 nm. The olive oil core is surrounded by a polymeric layer that consists of Epikuron 145V (commercial solution of phospholipids, Cargill Spain, Barcelona, Spain), Pluronic® F68 (non-ionic surfactant), and deoxycholic acid, which allows for the covalent bonding of IgG-αHER2 using the carbodiimide method (ECDI) [22]. This monoclonal antibody recognizes receptor HER2, which is overexpressed in some BC cell lines. We simulated the formation of the PC in vitro by incubating the NPs in the adequate medium supplemented simultaneously with fetal bovine serum (FBS) and fibrinogen (FB). We carried out a complete physico-chemical characterization of the LLNCs before and after functionalization, as well as once the PC was formed. This characterization allows us to verify the following: (i) the NPs are colloidally stable, (ii) the antibody is on the surface after functionalization, and (iii) the PC has been formed around the NPs. Through confocal microscopy and flow cytometry, we proved that LLNCs loaded with Nile Red (NR) can recognize the HER2 receptor in a BC cell line (SKBR-3) that overexpresses HER2 markers compared with HDFa, which does not express HER2, and how the formation of PC affects this specific recognition.

## 2. Results and Discussion

*2.1. Formulations and Physico-Chemical Characterization of Nanocarriers*

2.1.1. Preparation of Liquid Lipid Nanocapsules (LLNCs)

Nanocapsules were formulated using a previously reported slightly modified solvent-displacement technique [5,22,23], in which the only organic solvent used was ethanol in order to reduce the toxicity of the nanocapsules, and the aqueous phase was added to the organic phase in an abrupt way with high mechanical energy. The protocol is depicted schematically in Figure 2. After mechanical mixing of the aqueous and organic phases, a lipid liquid nanoemulsion is produced. This nanoemulsion contains nanocapsules, which are nanodroplets of olive oil stabilized in an aqueous medium by a complex shell formed by several components. These components, one of them being the non-ionic surfactant Pluronic F68, provide colloidal stability and enable convenient functionalization to enhance the LLNCs' applicability. Specifically, the NC features the following: (i) an ordered phospholipid monolayer, contributing to a negative surface electric charge [24]; (ii) hydrophilic chains from the poloxamer surfactant to improve the half life time [24,25]; and (iii) specific carboxylic surface groups from deoxycholic acid to allow for the covalent bonding of some specific molecules as antibodies [5]. The integration of these entire components in the nanocapsules was previously verified by nuclear magnetic resonance (NMR) [22].

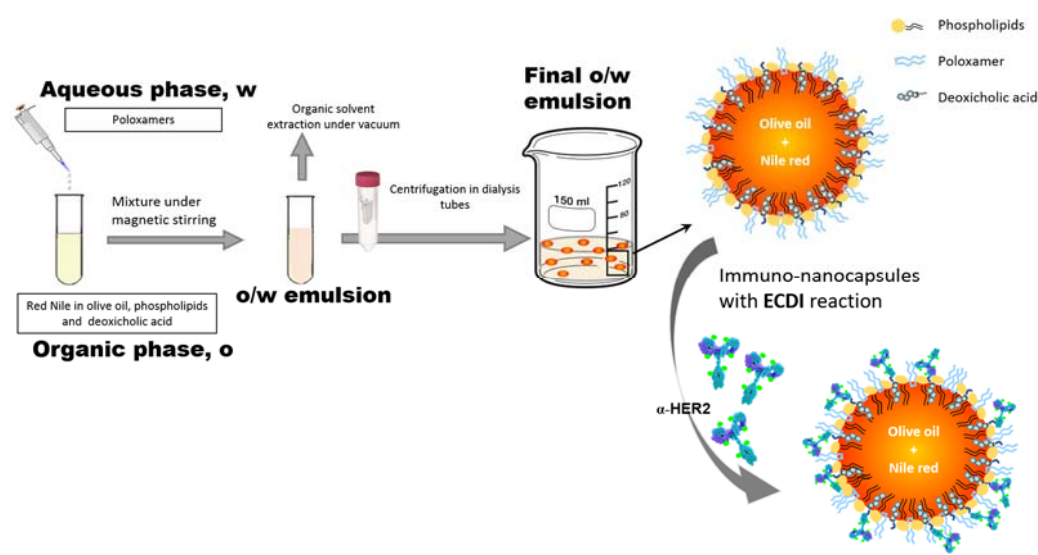

**Figure 2.** Schematic representation of the procedure steps for the preparation of LLCNs and subsequent functionalization to obtain immune-nanocapsules (LLNCs-$\alpha$HER2).

We verified that the protocol used in the synthesis of the nanocapsules resulted in colloidal systems with the necessary size and stability for subsequent biomedical applications by performing DLS hydrodynamic size measurements in pH 7.4 buffer (Table 1). After 1 month of 4 °C storage, these measurements were repeated, and no significant differences were detected. These results confirmed that these LLNCs have the appropriate diameter and stability for the purposes for which they were designed [26,27].

**Table 1.** Mean diameter, standard deviation (SD), and mode of the LLNCs, LLNCs-$\alpha$HER2, and LLNCs-$\alpha$HER2-PC measured at 25 °C with NTA and DLS techniques in pH 7.4 buffer.

| | NTA | | | DLS | |
|---|---|---|---|---|---|
| Sample | Mean Diameter (nm) | SD (nm) | Mode (nm) | Mean Diameter (nm) | PDI |
| LLNCs | 150 | 50 | 140 | 130 ± 20 | 0.14 ± 0.01 |
| LLNCs-$\alpha$HER2 | 150 | 50 | 151 | 140 ± 20 | 0.13 ± 0.02 |
| LLNCs-$\alpha$HER2-PC | 170 | 50 | 154 | 150 ± 30 | 0.16 ± 0.02 |

NTA was used as a complementary technique, which in addition to providing an average hydrodynamic diameter, allowed us to obtain the size distribution (Figure 3) and an estimation of the concentration of particles in the sample. This concentration was used in cellular assays.

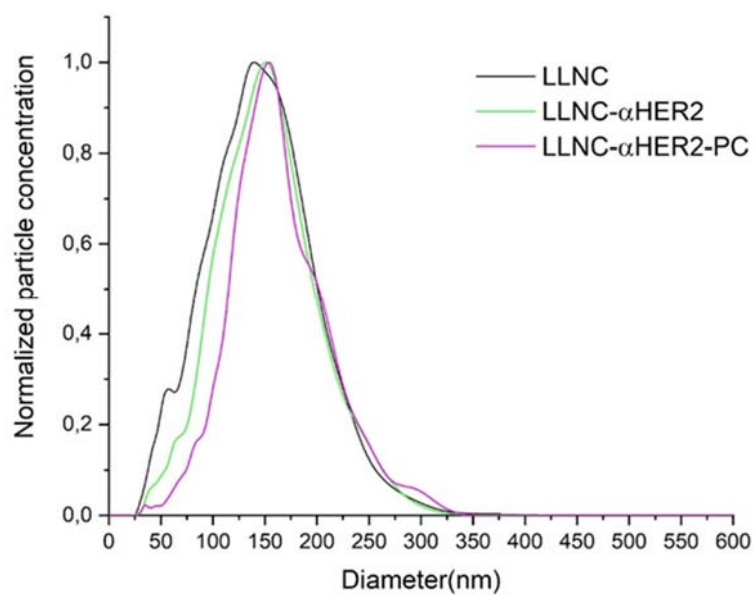

**Figure 3.** Hydrodynamic size distribution of the LLNCs, LLNCs-$\alpha$HER2, and LLNCs-$\alpha$HER2-PC measured at 25 °C with NTA technique in pH 7 buffer.

Figure 3 shows the size distribution with a low polydispersity and with a main peak (mode) of 140 nm. The mean diameter (Table 1) agrees with the value obtained by DLS and the polydispersity index value corresponds to a monodisperse nanosystem with a narrow diameter distribution. This result is also in accordance with those previously obtained following a similar formulation [22]. Finally, the electrokinetic behavior for LLNCs as a function of the medium pH reflects the chemical nature of weakly acidic surface charged groups, showing a reduction in the negative zeta potential value for pH below the p$K_a$ of phosphatidic and carboxylic groups [18] (Figure 4A).

2.1.2. Preparation of Liquid Lipid Immune-Nanocapsules (LLNCs-$\alpha$HER2)

It is essential to have nano-surfaces with specific biological properties in order to assess how the PC affects cellular interactions in a functionalized nanosystem. In this way, we chose a typical strategy using specific monoclonal antibodies and selected HER2 as the target membrane receptor, with a clear therapeutic application. In this way, we produced olive oil immuno-NCs with a defined amount of anti-HER2-FITC ($\alpha$-HER2) antibody. The protein concentration of the commercially available samples for this antibody limits the initial amount of protein incubated to 0.2 mg m$^{-2}$. This situation corresponds to a low coverage degree, which can be an advantage to reach adequate recognition and interaction with membrane receptors, allowing for an efficient and specific cell uptake [22] and the simultaneous location of fluorescent antibodies in cellular experiments. An excessive density of the molecules on the surface of the NPs can decrease the affinity of an antibody for their specific membrane substrate [28,29]. The EDCI covalent coupling protocol is routinely used for immobilization by covalent bonds of protein molecules (immunoglobulin G) on the surface of the LLNCs through different chemical groups (in this case, the carboxyl group provided by deoxycholic acid), and pH plays an important role, with it being necessary to adjust its value considering the isoelectric point (IEP) of the protein molecules [5]. In this case, the protocol previously optimized for different antibody molecules has been adapted to the specific characteristics of the α-HER2 antibody (IEP of 8.6), as it is described in detail in the Materials and Methods section. Thus, the pH of the reaction medium is adapted to reach a negative net charge regarding the electrical state of the antibodies, a condition that facilitates the effective union of these molecules through the carbodiimide method. This protocol does not guarantee a perfectly ordered spatial arrangement of the antibody molecules. However, a fraction of these is adequately arranged and a specific immunological recognition was previously contrasted when this covalent coupling protocol was used with similar LLNCs (see Figure 2). Thus, while the EDCI procedure yields satisfactory results for immunoreactions with specific antigens, the surface physical adsorption of antibody molecules in the absence of EDCI leads to immune-nanocapsules that do not show any specific immuno-agglutination response [5,22]. No aggregation of the LLNCs was detected during the coupling protocol. This point is crucial because it is necessary to prevent an uncontrolled increase in the nanosystem's size in order to preserve the capacity for an adequate bio-distribution for in vivo applications. After the dialysis step for cleaning, the first elution volume was analyzed by a spectrophotometric assay showing no presence of protein. This situation was previously described working with other specific antibodies for a similar low coverage degree, for which nearly total coupling was achieved, and the size of the LLNCs-αHER2 at pH 7.4 and the low-ionic strength medium remained similar to those of the original LLNC system [22]. DLS and NTA measurements did not show significant differences between bare and functionalized nanocapsules (Table 1), as is expected regarding the low coverage and the antibody molecule dimensions. Electrokinetic behavior and chromatography experiments for LLNCs-αHER2 corroborated the coupling of αHER2 molecules on the LLNC's surface (Figure 4).

It is widely described that the presence of proteins at the surface of colloidal particles produces a modulation of the surface electric charge, closing the IEP of the complex to the specific IEP of each protein. The protein coverage is decisive and the higher the degree of coating, the greater the ability to alter the original surface charge [30–32]. The net charge of protein molecules, positive under IEP, has the capacity to screen the negative surface charge of bare nanocapsules. As shown in Figure 4A, there are variations in the zeta potential data of LLNCs-αHER2 with respect to LLNCs for all pHs, demonstrating that protein immobilization on the surface of the nanocapsules was successful. In this case, the low coverage degree of the antibody molecules partially modified the original surface charge, reducing the absolute zeta potential value of the LLNCs. Even so, from neutral to basic pHs, the zeta potential for LLNCs-αHER2 kept a sufficient negative value, around −30 mV, to prevent colloidal aggregation by an electrostatic repulsion mechanism [33]. This situation correlates with DLS and NTA (Table 1 and Figure 3) results for LLNCs-αHER2, where the PDI value (<0.15) and the narrow size distribution correspond to colloidally stable nanosystems. The coupling of the αHER2 antibody was finally confirmed by SDS-PAGE. As can be seen in Figure 4B, the free αHER2 lane (lane 1) shows the two characteristic bands at 25 and 50 kDa, corresponding to the molecular weight of the Fab fraction and Fc moieties of immunoglobulins. The same bands were found for the LLNCs-αHER2 sample (lane 4), while the sample corresponding to the elution volume from the coupling experiment (lane 5) reflects the absence of antibody bands.

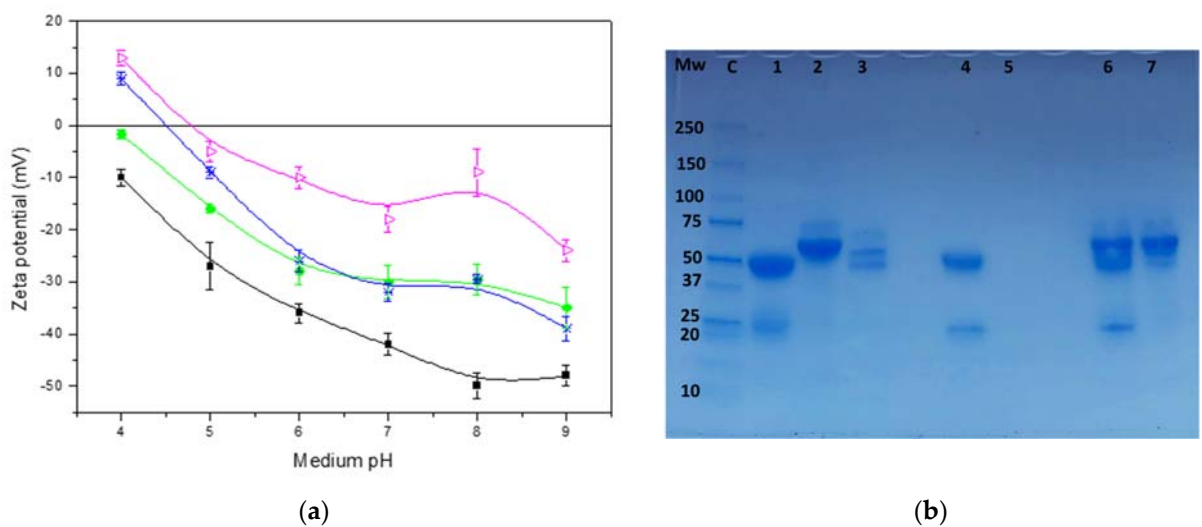

(**a**) (**b**)

**Figure 4.** Physico-chemical characterization: (**a**) Zeta potential of the LLNCs (■), LLNCs-αHER2 (●), LLNCs-αHER2-FBS (✶), and LLNCs-αHER2-PC (▷) measured at 25 °C as a function of the medium pH and low ionic strength. (**b**) SDS-PAGE analysis under reducing conditions of different LLNCs. (C) Molecular weight marker: (1) αHER2; (2) FBS; (3) Fibrinogen; (4) LLNCs-αHER2; (5) elution volume after cleaning LLNCs-αHER2; (6) LLNCs-αHER2-PC; (7) elution volume after cleaning LLNCs-αHER2-PC.

2.1.3. Preparation of Liquid Lipid Immune-Nanocapsules with a Protein Corona (LLNCs-αHER2-PC)

Once immune-nanocapsules (LLNCs-αHER2) were obtained, we carried out the in vitro formation of a PC surrounding the surface of LLNCs-αHER2. For this purpose, we used FBS, in which the globulin content is very low and albumin constitutes almost 70% of the total proteins [34,35], and FB (αα, ββ, and γγ chains), because albumin is the most abundant protein in blood, while FB is one of the main components of plasma [36]. Furthermore, both proteins are two of the most representative proteins found in hard and soft coronas, shaped in different lipidic or polymeric nanoparticles [37–39]. Bovine albumin (BSA) is normally used as a substitute for human albumin [40], and its use in PCs in in vitro experiments is widely described, being chosen as a test protein due to its availability, high stability, and solubility in water [41]. The use of a more complex biological fluid containing BSA, that is FBS, to mimic the biological environment to which these nanocapsules are exposed is also commonly described [25,42,43].

Thereby, after the incubation of the LLNCs-αHER2 in DMEM with 10% FBS supplemented with FB at a physiological plasma concentration (3 mg/mL) at 37 °C for 2 h with stirring, we obtained the LLNCs-αHER2-PC. Proteins show a strong trend to accumulate at interfaces, and blood proteins are strongly attracted by the NP surfaces [10]. The singular structure of LLNCs-αHER2 could strongly influence the adsorption of BSA and FB due to its high heterogeneity, merging phospholipid polar heads, hydrophilic chains from a poloxamer, carboxylic groups, and a low coverage of covalently attached antibody molecules.

Changes in the physico-chemical properties of the LLNCs-αHER2 were observed after the in vitro PC formation, which indicates that the surface has been modified again. Mean hydrodynamic diameter (both by DLS and NTA) increased (Table 1) due to the PC thickness. This experimental increase could be compatible with the presence of an additional protein cargo on the surface. A monolayer of albumin corresponds to a thickness between 3 and 4 nm [44], while the FB molecule has been characterized as a prolate ellipsoid of 47 nm × 10.5 nm [45]. The interaction of FB with lipidic nanoparticles of different surface components has been described, where the greater the surface hydrophobicity, the less protein coating, which corresponds to a side-on orientation of this protein on the surface. Furthermore, a low value of PDI was maintained despite the particle size increment [46].

Size values from Table 1 for LLNCs-αHER2-PC show a very similar tendency with an increase that could correspond to the presence of both proteins, BSA and FB, structured in an irregular or discontinuous monolayer preserving the colloidal stability at physiological pH (Figure 5).

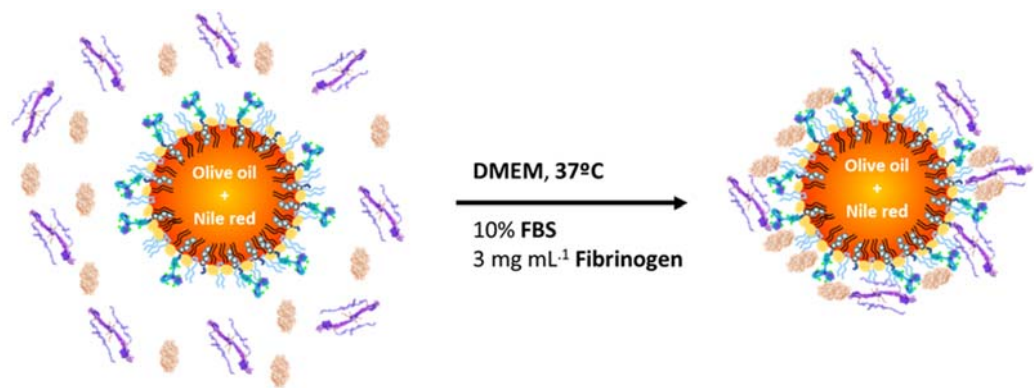

**Figure 5.** LLNCs-αHER2-PC after incubation of immune-nanocapsules in a simulated physiological medium with FBS supplemented with FB.

Faizullin et al. additionally described a shifting of the zeta potential to less negative values due to the presence of protein charge patches [46]. Indeed, this is the situation that we have observed with LLNCs-αHER2-PC complexes. DLS experiments reflected the presence of albumin (BSA) and FB in LLNCs-αHER2 after in vitro incubation of immune-nanocapsules to compose a surface protein corona. As can be seen in Figure 4A, the electrokinetic behavior as a function of the medium pH, expressed in terms of the zeta potential, results in a modification that shows the influence of surface proteins. This parameter becomes less negative because of the proteins forming the corona. In order to differentiate the contribution of each PC protein, the electrokinetic tendency of an additional complex incubated only with FBS, LLNCs-αHER2-FBS, was analyzed. Results from Figure 4A confirm the presence of BSA and FB on the surface of the LLNCs-αHER2-PC nanosystem. Firstly, the complex incubated with only FBS shows the typical electrokinetic behavior of colloidal nanoparticles covered by BSA, and the acidic IEPs of albumin (pI = 4.7) even promote a change to positive zeta potential values at pH 4. As has been previously commented, the IEP of LLNC–protein complexes gradually tends towards those of proteins loaded on the surface according to the coverage degree because of the partial or total screen of original LLNCs' surface charge [32,33,47].

Secondly, for LLNCs-αHER2-PC, a mixture of both proteins seems to be clear and the influence of FB slightly shifts the IEP of the complexes, bringing it closer to those corresponding to FB chains (pI between 5.1 and 6.3 depending on the chain) [48]. At the same time, the net zeta potential value is reduced at neutral and basic pHs, which has been described for proteins such as IgG and FB [31,33]. In fact, it has been described how FB induces the colloidal aggregation of NPs in a protein concentration-dependent way [49]. This is consistent with the reduction in zeta potential and the partial or total screen of electrostatic repulsion, one of the interactions that must prevent the colloidal destabilization. In our case, for LLNCs-αHER2-PC, the equilibrate mixture of BSA and FB on the surface, beside the possibility of additional stabilization mechanisms due to steric and/or hydration forces previously described for this type of nanocapsule [22], would be enough to prevent aggregation in consonance with DLS and NTA measurements.

Finally, the SDS-PAGE experiment from Figure 4B eventually confirmed the presence of BSA and FB in the LLNCs-αHER2-PC. Characteristic bands of BSA (67 kDa) and FB chains (66.2, 54.5, and 48.4 kDa) of lanes 2 and 3 were reproduced in lane 6 for LLNCs-

αHER2-PC and additionally in lane 7 corresponding to the elution volume of the first cleaning step after the PC conformation experiment. This result reflects that only a fraction of both proteins remains at the surface as a consequence of the initial concentration excess and the limited affinity of these protein molecules for an LLNC's surface designed to limit these kinds of interactions. Taking into account the absence of proteins after the following cleaning steps, we could consider that this protein fraction presents a firm attachment to the surface, constituting a hard PC.

It has been previously shown how the presence of hydrophilic polymers as polyethylene glycols or poloxamers on the surface significantly decreases the protein adsorption on different types of nanoparticles, being able to prevent or diminish the formation of the hard corona [25,50,51]. Additionally, it has been recently described how the disposition of stable artificial coronas, for example, using antibody molecules (IgG), could aid in preserving stealth properties regulating cellular interactions of nanosystems in physiological mediums [13]. An adequate functionalization, controlling the surface charge, chemistry, and roughness makes it possible to control the PC formation [43]. Even more, a control of pre-coatings with different protein molecules such as IgGs, albumins, Apo, or FB could improve properties of nano-delivery systems including targeting efficacy or blood circulation time [10].

Concluding this section, the LLNCs-αHER2-PC nanosystem has the surface structure and composition to prove how some representative blood proteins could affect the cell interactions and targeting properties of a specific functionalized immune-nanocarrier such as LLNCs-αHER2.

*2.2. Specific Cellular Uptake of Immuno-Nanocapsules*

The olive oil cores of all our nanosystems were labeled beforehand with the fluorophore Nile Red (NR) to investigate the cellular entry as previously published by us, showing the presence of this fluorescent molecule in the hydrophobic oil core of our LLNCs and its absence in the external aqueous medium [18]. Moreover, the anti-HER2 antibody bound to LLNCs was labeled with FITC as we mentioned before. These two fluorophores allowed us to track our LLNCs and to know the behavior of our LLNCs in in vitro experiments like confocal microscopy or flow cytometry. Thus, the cellular uptake of NR-LLNCs, NR-LLNCs-HER2, and NR-LLNCs-HER2-PC was investigated on SKBR3 (high expression of HER2) and HDFa (no expression of HER2) by both confocal microscopy and flow cytometry. The obtained confocal microscopy images are shown in Figures 6 and S1–S6.

First, we performed a characterization of the expression levels of HER2 on both cell lines using free HER2-FITC antibody. The fluorescence intensity obtained was 63,573.5 in SKBR3 (which could be assumed as 100% of fluorescence intensity) and 552,5 in HDFa (which could be assumed as 1.5% of fluorescence intensity) (Figure S7, "C+"). Negative control fluorescence intensity histograms are shown in Figure S7, "C−". These values were satisfactory and can thus conclude that the great difference in the expression levels of HER2 between SKBR3 and HDFa is adequate to corroborate this study. Confocal microscopy images show how NR-LLNCs enter into both types of cell populations after 60 min of accumulating NR in the cytoplasm (Figure 6A,B). Images with all three nanocapsules at all incubation times are displayed in Figures S1-S6. In the case of NR-LLNCs-HER2, where αHER2 was labeled with FITC, we observed the specific recognition of the HER2 membrane receptors by these LLNCs (Figure 6A, "Green filter") and the release into SKBR3 cells of NR after 60 min of incubation (Figure 6A, "Merge") contrary to what happens in HDFa (Figure 6B "Green filter"). These NR-LLNCs-HER2 penetrated the SKBR3 cells, which suggests a specific recognition of the HER2 receptors.

These results may suggest that our nanosystem enters into the cells by a clathrin-mediated endocytosis mechanism, which has been reported as the main mechanism occurring through surface receptors with nanoparticles smaller than 200 nm [52,53]. In the case of NR-LLNCs-HER2-PC, we could also observe the specific surface interaction of the

αHER2 labeled with FITC but with an evident reduction in the recognition level of HER2 receptors in SKBR3 cells (Figures 6A, S2 and S3). These results suggest that the simulated PC formed around the immune-LLNCs does not block this specific recognition but reduces it by partially masking the αHER2 bond to the LLNCs. There are examples in which pre-adsorbed antibodies on the surface of NPs are integrated into the NP–corona complex after an incubation with human plasma, at the same time maintaining their targeting specificity. In this case, magnetosomes around 100 nm in diameter were functionalized with affibodies that attached to anti-HER2 humanized antibodies with an adequate surface disposition [54].

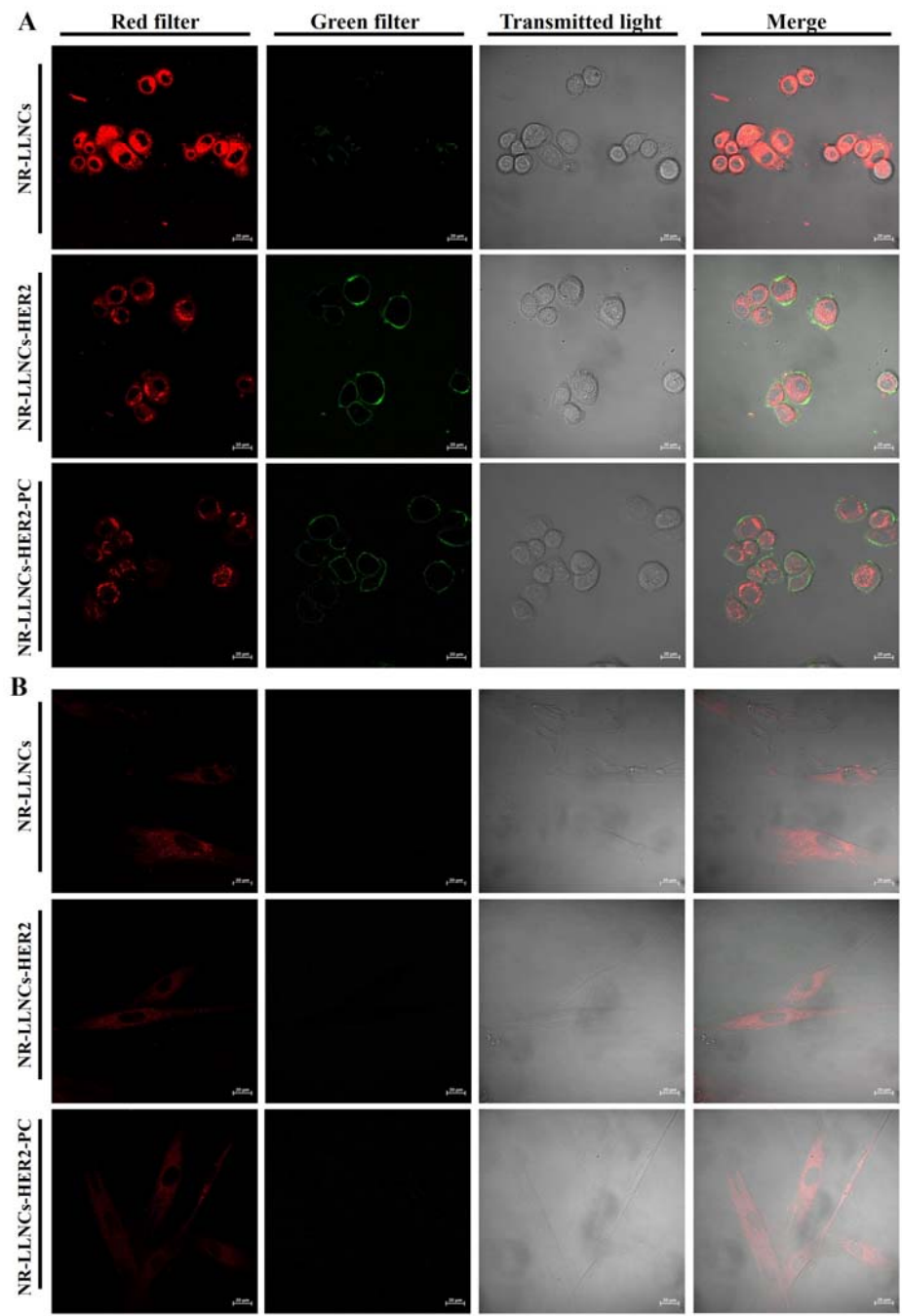

**Figure 6.** Representative confocal microscopy of (**A**) SKBR3 and (**B**) HDFa (scale bar = 20 μm) incubated for 60 min with NR-LLNCs, NR-LLNCs-HER2, and NR-LLNCs-HER2-PC. Red filter and green filter correspond to LLNCs labeled with Nile Red and LLNCs labeled with HER2-FITC, respectively.

The flow cytometry assay confirmed the uptake and the specific recognition of our LLNC-based nanosystems in the two cell populations (Figure 7 and S9). In that graph, we represent the mixture of red and green fluorescence intensity belonging to cells that have our LLNC-based nanosystem inside them. The analysis showed the differential uptake efficiency and specificity for NR-LLNCs in SKBR3 and HDFa depending on the functionalization with the $\alpha$HER2 antibody. Moreover, because there were differences in the expression levels of the HER2 membrane receptor for both cell lines, we also found a different behavior for each type of LLNC. However, most bare NR-LLNCs entered into both cell lines regardless of their HER2 expression level (Figure S7). In contrast, for NR-LLNCs-HER2, the entry was dependent on the HER2 expression in the membrane of both cell populations, with uptake values in SKBR3 around 7387.5; 23,930.5; 37,190.5; and 47,166 (whose normalized values are 0.14513, 0.50065, 0.78562, and 1) at 0, 15, 30, and 60 min, respectively, and only around 634.5, 675.5, 675.5, and 661 (whose normalized values are practically 0) at 0, 15, 30, and 60 min, respectively, in HDFa (Figure 7). Representative dot plots are shown in Figure S8. This suggests that NR-LLNCs-HER2 actively recognizes the HER2 receptor overexpressed on SKBR3, producing the cell internalization of the nanosystem, contrary to what happens with HDFa, where these LLNCs displayed practically null internalization (50-fold reduction). These results are in concordance with previous studies showing the targeting cells overexpressing a specific surface marker by LLNCs functionalized with a specific antibody [22,55,56]. It is worth highlighting the selective uptake efficiency of NR-LLNCs-HER2 in SKBR3 BC as opposed to human HDFa, which is a very similar result to those described by Trabulo et al. using an anti-CD47 antibody covalently attached to the surface of iron oxide nanoparticles for the recognition of this membrane receptor overexpressed in pancreatic cancer cells [57].

Moreover, because the interaction of HER2 promotes EGFR-mediated pathways, consequently leading to tumor cell growth, tumor cell migration, and chemotherapy resistance in solid cancers [58], the ability of LLNCs functionalized with the $\alpha$HER2 antibody could bind and neutralize the receptor by competitive inhibition of its HER2 ligand and consequently prevent the receptor-signaling cascade activation. All this makes this the designed nanosystem a good candidate for targeted therapy against BC. On the other hand, one of the main goals of this work, as mentioned before, was to evaluate the influence of the PC formation onto the nanoparticle surface in the recognition of cell surface markers. In the same manner, we can observe how NR-LLNCs-HER2 penetrated the SKBR3 cells as time went by, while NR-LLNCs-HER2-PC reduced their uptake by 40% (Figure 7). Su et al. have reported, working with colloidal gold NPs functionalized with specific integrin peptides, that the targeting abilities of cRGD were maintained despite the formation of a surface PC, although the cellular uptakes were decreased by around 40%. This indicated that the binding efficiencies between cRGD molecules and their receptors were decreased [59].

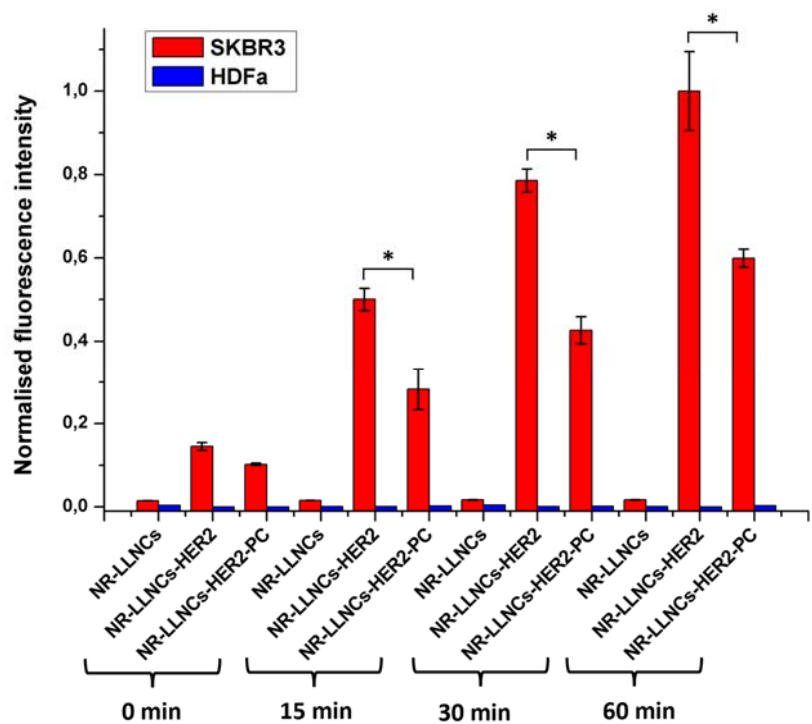

**Figure 7.** In vitro uptake of NR-LLNCs, NR-LLNCs-HER2, and NR-LLNCs-HER2-PC in both SKBR3 and HDFa analyzed by flow cytometry. Values represent the fluorescence intensity of the merge channels for green and red as NR-LLNCs-HER2 are both red and green fluorescent. Data are mean values ± SD. * $p$ < 0.001 shows the significant values calculated using *t*-test.

These results showed that the simulated PC is partially blocking the active recognition of the HER2 surface marker by NR-LLNCs-*α*HER2. The characteristic surface composition combining hydrophilic polymeric tails with phospholipids and the directional covalent anchoring of specific antibody molecules must be enough to control the surface PC, preserving targeting capabilities and adequate specific therapeutic effect in spite of the reduction in the cellular uptake. This behavior has been contrasted working in vivo with an animal model using a nanocarrier with the same surface properties but using anti-CD44 receptor antibodies, where nanocarriers show a specific accumulation in the target tissues avoiding or reducing the MPS clearance [22]. Although several reviews have published the negative impact of PC on the behavior and targeting capability of nanoparticle-based delivery systems [60,61], there is an increasing tendency to show the use of different strategies that make it possible to regulate the PC formation, not only to avoid this surface structure but also to use it with the best interest looking for an equilibrium between targeting properties and physiological behavior [11,15,16,62].

## 3. Materials and Methods

### 3.1. Materials

Poloxamer 188 (Pluronic® F68), N-(3-dimethylaminopropyl)-N'-ethyl-carbodiimide (ECDI), RPMI-1640 Medium, olive oil, Nile Red (NR), deoxycholic acid (DC), fetal bovine serum 10% (FBS), *β*-mercapto-ethanol, fibrinogen from human plasma 50–70% (FB), sodium dodecyl sulphate (SDS), tris(hydroxymethyl)aminomethane (TRIS), and Brillant Blue R-250 were purchased from Sigma-Aldrich (Madrid, Spain). All of them, except the olive oil, were used as received. Olive oil was previously purified in our laboratories with activated magnesium silicate (Florisil, Fluka) to eliminate free fatty acids. Mini-PROTEAN TGX Stain-Free Precast Gels and a Precision Plus ProteinTM Standards Kaleidoscope were supplied from Bio-Rad (Madrid, Spain). Epirukon 145V, which is a phosphatidylcholine-enriched fraction of soybean lecithin, was from Cargill (Barcelona, Spain). The monoclonal

ErbB-2 antibody anti-human Vio®Bright FITC (αHER2) (clone 24D2) from humans was obtained by Miltenyi Biotech (Madrid, Spain). Its isoelectric point (IEP) is 8.6 [63]. DMEM (Dulbecco's Modified Eagle Medium) with no phenol red was obtained from Gibco™ (Waltham, MA, USA). Human Fibroblast Expansion Medium (HFEM) was purchased from ThermoFisher scientific (Waltham, MA, USA). The water was purified in a Mili-Q Academic Millipore system. Other solvents and chemicals used were of the highest commercially available grade.

### 3.2. Cell Lines and Culture Conditions

The SKBR3 human breast cancer (BC) cell line was obtained from American Type Culture Collection (ATCC) and cultured following ATCC recommendations. The Human Dermal Fibroblasts (HDFa) cell line was purchased from ThermoFisher scientific. HDFa was cultured in HFEM and the SKBR3 BC cell line was cultured in RPMI-1640 medium (RPMI) supplemented with 10% (*v*/*v*) heat-inactivated fetal bovine serum (FBS), 1% L-glutamine, 2.7% sodium bicarbonate, 1% Hepes buffer, and 1% penicillin/streptomycin solution (GPS, Sigma). Both cell lines were grown at 37 °C in an atmosphere containing 5% $CO_2$. All cell cultures tested negative for mycoplasma infection.

### 3.3. Preparation of Lipid Nanocapsules

The synthesis of the olive oil Lipid Liquid Nanocapsules (LLNCs) was carried out following the principle of the solvent displacement technique using a slight modification of a method previously described [5,22].

Briefly, 20 mL of an aqueous phase under magnetic stirring formed by 50 mg of Pluronic® F68 was produced. An organic phase consisting of 125 μL of olive oil, 40 mg of Epikuron 145V, and 10 mg of deoxycholic acid (provides the -COOH groups to make the vectorization with antibody molecules possible) was dissolved by sonication in 10 mL of ethanol (98% synthetic grade) and strongly added to the first aqueous phase under magnetic stirring. The mixture turned milky immediately due to the formation of a nanoemulsion. The organic solvent, ethanol, plus a portion of the volume of water were evaporated in a rotary evaporator at 35 °C, giving a final volume of 16 mL. The sample was then cleaned of surfactant residues and other substances by borate buffer dialysis (1.13 mM and pH 10) with a 300 KDa membrane under magnetic stirring for 24 h at 4 °C. Additionally, NR fluorochrome was previously dissolved in the olive oil phase in order to synthesize nanocapsules loaded with these compounds inside.

### 3.4. Preparation of Antibody-Coated Nanocapsules

The functionalization of the surface of LLNCs with αHER2 antibody molecules to transform them into immune-nanocapsules was carried out by following the protocol of the carbodiimide method (ECDI) with the needed modifications [5]. The previous dialysis step of LLNCs using an activation buffer with a low ionic strength and pH 10 allows for the adaptation of the medium to the IEP of αHER2 molecules, reaching the best efficiency of surface antibody immobilization. The methodology of this functionalization consists, initially, in the addition of 1 mL of a borate buffer solution (pH 10.0) of ECDI at 15 mg/mL to the LLNC solution having a total particle surface equal to 0.29 $m^2$ for the covalent binding of αHER2. Subsequently, 6 μL of β-mercaptoethanol was added, which inactivated EDCI that had not bonded to the surface of the nanocapsules. Next, the antibody coverage was performed by adding α-HER at a concentration of 0.2 mg/$m^2$, and then the solution was incubated in the dark at room temperature for 2 h. Finally, the LLNCs-αHER underwent three centrifugation cycles using Vivaspin 20 centrifugal tubes MwCO 1000 kDa (Sartorius, Göttingen, Germany) at 5000 rpm for 15 min to remove any antibody molecules that were not coupled to the nanocapsules' surface. The first elution volume was collected for protein quantification by ultraviolet spectrophotometry and BCA (bicinchoninic acid assay) methods. The final LLNCs-αHER2 stock suspended in a phosphate buffer at pH 7.4 was stored at 4 °C for further use.

### 3.5. In Vitro Protein Corona Formation

Two protein adsorption assays were performed on covalently functionalized lipid nanocapsules, LLNCs-αHER2. Firstly, 1 mL of LLNCs-αHER2 was mixed with 4 mL of DMEM. Previously, this medium was supplemented with 10% FBS, obtaining the system LLNCs-αHER2-FBS in this way. In the second experiment, DMEM was supplemented with 10% FBS and 3 mg/mL of FB and it was labeled as LLNCs-αHER2-PC.

All these samples were incubated with agitation for 2 h at 37 °C. Finally, they were cleaned in centrifuge tubes with a membrane with a pore size of 1000 kDa at 5000 rpm, performing the necessary centrifugation cycles to guarantee the cleaning of nanocapsules. In each cycle, the sample volume was reduced to 1 mL and the suspension medium was renewed by adding a phosphate buffer of pH 7.4. After the washing protocol, LLNCs-αHER2-PC remained in the aqueous dispersion in absence of free proteins.

### 3.6. Protein Separation by SDS-PAGE

In order to check the bio-molecular composition of the LLNC's surface after the covalent functionalization with αHER2 and the subsequent formation of the PC, different SDS-PAGE (Poly-Acrylamide Gel Electrophoresis) tests were performed. Samples including protein controls, immune-nanocapsule complexes, and elution volume samples from cleaning procedures were treated and denatured by boiling 10 μL of each sample for 5 min at 95 °C in the following buffer (Laemmli buffer) [64]: 62.5 mM Tris-HCl (pH 6.8 at 25 °C), 2% (*w/v*) sodium dodecyl sulfate (SDS), 10% glycerol, 0.01% (*w/v*) bromophenol blue, and 40 mM dithiothreitol (DTT). Next, proteins were separated depending only on their molecular weight [65] into a Mini-PROTEAN TGX Stain-Free Precast with 4–15% of porous polyacrylamide gel under denaturing conditions using Sodium-Dodecyl-Sulphate (SDS) by using an electrical potential of 150 V for 70 min. After that time, the gels were left in Blue Coomassie (2.5 g of Coomassie Brilliant Blue R-250, 450 mL of methanol, 100 mL of acetic acid, and 450 mL of ultrapure water) with agitation for 15 min. Finally, the gels were washed 3 times with distilled water and 3 times with a destaining solution (9:9:2, methanol:ultrapure water:acetic acid) for 30 min each time. The gel was then left in distilled water overnight.

### 3.7. Physico-Chemical Characterization of LLNCs, LLNCs-HER2, and LLNCs-HER2-PC: Hydrodynamic Size Distribution, Electrokinetic Behavior, and Nanocapsule Concentration

The hydrodynamic size distribution and the polydispersity index (PDI) were measured using dynamic light scattering (DLS) with a Zetasizer Nano Zeta ZS, Malvern Instruments, Malvern, UK. The light scattered by the samples was detected at 173°, and the temperature was set at 25 °C. The diffusion coefficient measured by dynamic light scattering can be used to calculate the size of the different nanosystems by means of the Stokes−Einstein equation. The polydispersity index (PDI) calculated by the analysis of the intensity autocorrelation function informs about the homogeneity of the size distribution [66].

The surface charge of different nanosystems was estimated by means of the electrophoretic mobility (which allows us to calculate the zeta potential) of the nanocapsules using a Doppler Laser Electrophoresis (DLE) technique, which is based on the frequency shift of a laser beam that strikes a scattering of charged particles in motion due to the application of an oscillating external electric field. This parameter was also measured with the same DLS device at 25 °C in different media with low ionic strength and a pH ranging from 4 to 10.

In addition, the hydrodynamic size distribution was measured and assessed using Nanoparticle Tracking Analysis (NTA) with a NanoSight LM10-HS(GB) FT14 (NanoSight, Amesbury, UK) and a sCMOS camera. From at least three measures of size distribution (concentration of particles versus diameter) at a Camera Level of 14, the concentration of nanocarriers was also calculated.

*3.8. Uptake Studies of LLNCs, LLNCs-HER2, and LLNCs-HER2-PC by Confocal Microscopy and Flow Cytometry*

Cells ($1.5 \times 10^5$) from both SKBR3 and HDFa were seeded into Corning cell culture T75 flasks (Corning, Madrid, Spain). In this study, LLNCs were formulated by adding NR into olive oil at a concentration of 0.025% (*w/w*). Regarding NR, we assumed that because of its lipophilic nature, it would be fully incorporated into the olive oil core of the LNNCs as previously verified [18]. In any case, we performed an assay to evaluate the possible exclusion of Nile Red from the LLNCs. Briefly, several aliquots of NR-LLNCs were incubated at 37° C for 4, 8, and 24 h in phosphate buffer (PB), PB saline (PBS), and phenol red-free DMEM with FBS [22]. These samples were centrifuged in Vivaspin tubes (1000 kDa pore size), and supernatants were collected and analyzed using a spectrofluorometer. The membrane binding and cell internalization of three labeled LLNCs, NR-LLNCs, NR-LLNCs-αHER2, and NR-LLNCs-αHER2-PC, were examined by laser-scanning confocal microscopy. Briefly, the two types of cell populations ($3 \times 10^3$/well) were seeded in Slide 8-Well chambers (ibiTreat, IBIDI) with RPMI medium and HFEM, respectively. After 48 h, the required concentration of the three labeled LLNCs was added to the cells. The images were taken at 0, 1, 5, 10, 15, 30, and 60 min of incubation. Imaging experiments were conducted with a Zeiss LSM 710 laser-scanning microscope (Madrid, Spain) using a tissue culture chamber (5% $CO_2$, 37 °C) with a plan-apochromat 63×/1.40 Oil DIC m27. The Argon laser was used at 488 nm at 2% potency (red) (max. potency Argon laser 25 mW). The HeNe laser was used at 543 nm at 22% potency (green) (max. potency HeNe laser 1.2 mW). Images were processed with Zen ILite 2.1 software. Additionally, a flow cytometry assay was performed. Firstly, we performed a characterization of SKBR3 and HDFa in order to corroborate the HER2 expression levels. Briefly, the cell surface marker levels of SKBR3 and HDFa were determined with human antibody αHER2-FITC (ErbB-2 antibody anti-human Vio®Bright FITC). Samples were measured and analyzed by flow cytometry on a FACS CANTO II (BD Biosciences, Franklin Lakes, NJ, USA). For the uptake study, once the different cell populations were obtained, the two cell populations were washed and centrifuged at 1500 rpm for 5 min in tubes and resuspended in a 1% BSA (bovine serum albumin) PBS solution to block the possible nonspecific binding of antibodies. Next, $1 \times 10^6$ cells per sample were incubated with the required concentration of the three labeled LLNCs at 37 °C. The incubation times of the cells with the LLNCs were 0, 15, 30, and 60 min. Then, the cells were washed with 1× PBS and centrifuged at 1500× *g* for 5 min twice to remove the non-internalized LLNCs. Finally, the cells were resuspended in 300 μL of 1× PBS and analyzed for red and green fluorescence by flow cytometry (FACS CANTO II (BD Biosciences)) using FACSDiva™ v9.0 software. Laser lines: 488 nm (blue) and 633 nm (red). Filters: for blue laser, 502LP-530/30BP (FITC); 556LP-582/42BP (PE); 610LP, 655LP 670LP (PerCP Cy5.5); and 735LP-780/60BP (PE-Cy7); for red laser, 660/20BP (APC) and 735LP-780/60BP (APC-Cy7). All experiments were performed in triplicate and replicated at least twice. Sterility evaluations of all nanosystems were performed prior to developing LLNC uptake studies in order to exclude possible biological contamination. Cells treated with naked LLNCs were used as negative controls.

*3.9. Statistical Analysis*

The data are presented as the mean ± the standard deviation in the error bars. The sample size (n, *n*) indicates the experimental repeats of a single representative experiment, being 3 unless otherwise specified. The results of the experiments were validated by independent repetitions. Graphs and statistical difference data were made with GraphPad

Prism 6.0 (Graphpad Software Inc., La Jolla, CA, USA). Statistical significance was determined using Student's *t*-test in paired groups of samples with a known median. A *p*-value of ≤0.01 was considered significant.

## 4. Conclusions

In summary, we have used a versatile nanosystem with the ability to be efficiently coated with fluorescent antibody molecules, obtaining, in this way, a successful functionalization for a specific immunological recognition of the HER2 cell membrane receptors, which can be monitored by using fluorescent techniques. Additionally, the oil core charges a typical fluorophore and the hydrophilic surface chains of the LLNCs-$\alpha$HER2 contribute to improve the stealth behavior under physiological conditions. Using this multifunctionalized nanocarrier, we have composed on its surface a PC with representative serum and plasma proteins. A complete physico-chemical characterization shows that our nanocapsules, LLNCs-$\alpha$HER2 and LLNCs-$\alpha$HER2-PC, have the adequate colloidal properties and the surface protein composition in order to study the impact of the PC on the targeting abilities of these nanocarriers. Furthermore, the specific immunological recognition of the HER2 cell membrane receptors has been revealed in vitro, showing the adequate surface disposition of the HER2 antibody molecules immobilized on the LLNCs. Our in vitro results also supported the enhanced targeting activity and receptor-mediated binding mechanism of LLNCs-$\alpha$HER2 with HER2 overexpressing BC cells. Finally, our results show how the surface characteristics of the immune-nanocapsules could modulate the formation of a PC surrounding the nanocarrier without avoiding the specific recognition of HER2 receptors but decreasing the uptake efficiency of the cells. Taking that into account, it has been corroborated that the PC formation around a nanosystem is a key parameter to keep in mind for nanotechnology development.

**Supplementary Materials:** The following supporting information can be downloaded at www.mdpi.com/xxx/s1.

**Author Contributions:** Conceptualization, S.N.M., A.B.J.R., J.A.M. and J.M.P.G.; methodology, S.N.M., T.d.C.S. and P.G.; investigation, M.M.C., D.C.S., S.N.M., T.d.C.S., P.G., A.B.J.R. and J.M.P.G.; resources, A.B.J.R. and J.A.M.; writing—original draft preparation, S.N.M., A.B.J.R. and J.M.P.G.; writing—review and editing, S.N.M., P.G., T.d.C.S., A.B.J.R., J.A.M. and J.M.P.G.; supervision, A.B.J.R. and J.M.P.G.; project administration, A.B.J.R. and J.A.M.; funding acquisition, A.B.J.R. and J.A.M. All authors have read and agreed to the published version of the manuscript.

**Funding:** This research was funded by FEDER/Junta de Andalucía-Consejería de Transformación Económica, Industria, Conocimiento y Universidades, Projects PY20_00241 and A-FQM-90-UGR20. The authors thank MCIN/AEI/10.13039/501100011033/FEDER "Una manera de hacer Europa" for funding the PID2022-140151OB-C21 and PID2022-140151OB-C22 projects and the Chair "Doctors Galera-Requena in cancer stem cell research".

**Institutional Review Board Statement:** Not applicable.

**Informed Consent Statement:** Not applicable.

**Data Availability Statement:** Data are contained within the article or Supplementary Materials.

**Conflicts of Interest:** The authors declare no conflict of interest.

## References

1. Sechi, M.; Sanna, V.; Pala, N. Targeted therapy using nanotechnology: Focus on cancer. *Int. J. Nanomed.* **2014**, *2014*, 467–483. https://doi.org/10.2147/IJN.S36654.
2. Shi, J.; Kantoff, P.W.; Wooster, R.; Farokhzad, O.C. Cancer nanomedicine: Progress, challenges and opportunities. *Nat. Rev. Cancer* **2017**, *17*, 20–37. https://doi.org/10.1038/nrc.2016.108.
3. Xu, X.; Ho, W.; Zhang, X.; Bertrand, N.; Farokhzad, O. Cancer nanomedicine: From targeted delivery to combination therapy. *Trends Mol. Med.* **2015**, *21*, 223–232. https://doi.org/10.1016/j.molmed.2015.01.001.

4. Cirstouihapca, A.; Bossynobs, L.; Buchegger, F.; Gurny, R.; Delie, F. Differential tumor cell targeting of anti-HER2 (Herceptin®) and anti-CD20 (Mabthera®) coupled nanoparticles. *Int. J. Pharm.* **2007**, *331*, 190–196. https://doi.org/10.1016/j.ijpharm.2006.12.002.
5. Sánchez-Moreno, P.; Ortega-Vinuesa, J.L.; Boulaiz, H.; Marchal, J.A.; Peula-García, J.M. Synthesis and characterization of lipid immuno-nanocapsules for directed drug delivery: Selective antitumor activity against HER2 positive breast-cancer cells. *Biomacromolecules* **2013**, *14*, 4248–4259. https://doi.org/10.1021/bm401103t.
6. Dakshinamurthy, P.; Mukunda, P.; Prasad Kodaganti, B.; Shenoy, B.R.; Natarajan, B.; Maliwalave, A.; Halan, V.; Murugesan, S.; Maity, S. Charge variant analysis of proposed biosimilar to Trastuzumab. *Biologicals* **2017**, *46*, 46–56. https://doi.org/10.1016/j.biologicals.2016.12.006.
7. Farokhzad, O.C.; Langer, R. Impact of Nanotechnology on Drug Delivery. *ACS Nano* **2009**, *3*, 16–20. https://doi.org/10.1021/nn900002m.
8. Francia, V.; Yang, K.; Deville, S.; Reker-Smit, C.; Nelissen, I.; Salvati, A. Corona Composition Can Affect the Mechanisms Cells Use to Internalize Nanoparticles. *ACS Nano* **2019**, *13*, 11107–11121. https://doi.org/10.1021/acsnano.9b03824.
9. Pareek, V.; Bhargava, A.; Bhanot, V.; Gupta, R.; Jain, N.; Panwar, J. Formation and Characterization of Protein Corona Around Nanoparticles: A Review. *J. Nanosci. Nanotechnol.* **2018**, *18*, 6653–6670. https://doi.org/10.1166/jnn.2018.15766.
10. Singh, N.; Marets, C.; Boudon, J.; Millot, N.; Saviot, L.; Maurizi, L. In vivoprotein corona on nanoparticles: Does the control of all material parameters orient the biological behavior? *Nanoscale Adv.* **2021**, *3*, 1209–1229. https://doi.org/10.1039/d0na00863j.
11. Mirshafiee, V.; Kim, R.; Park, S.; Mahmoudi, M.; Kraft, M.L. Impact of protein pre-coating on the protein corona composition and nanoparticle cellular uptake. *Biomaterials* **2016**, *75*, 295–304. https://doi.org/10.1016/j.biomaterials.2015.10.019.
12. Lundqvist, M.; Augustsson, C.; Lilja, M.; Lundkvist, K.; Dahlb?ck, B.; Linse, S.; Cedervall, T. The nanoparticle protein corona formed in human blood or human blood fractions. *PLoS ONE* **2017**, *12*, e0175871. https://doi.org/10.1371/journal.pone.0175871.
13. Digiacomo, L.; Pozzi, D.; Palchetti, S.; Zingoni, A.; Caracciolo, G. Impact of the protein corona on nanomaterial immune response and targeting ability. *WIREs Nanomed. Nanobiotechnology* **2020**, *12*, e1615. https://doi.org/10.1002/wnan.1615.
14. Martín-Contreras, M.; Navarro-Marchal, S.A.; Peula-García, J.M.; Jódar-Reyes, A.B. Progress and Hurdles of Therapeutic Nanosystems against Cancer. *Pharmaceutics* **2022**, *14*, 388. https://doi.org/10.3390/pharmaceutics14020388.
15. Farshbaf, M.; Valizadeh, H.; Panahi, Y.; Fatahi, Y.; Chen, M.; Zarebkohan, A.; Gao, H. The impact of protein corona on the biological behavior of targeting nanomedicines. *Int. J. Pharm.* **2022**, *614*, 121458. https://doi.org/10.1016/j.ijpharm.2022.121458.
16. Khan, S.; Sharifi, M.; Gleghorn, J.P.; Babadaei, M.M.N.; Bloukh, S.H.; Edis, Z.; Amin, M.; Bai, Q.; Ten Hagen, T.L.M.; Falahati, M.; et al. Artificial engineering of the protein corona at bio-nano interfaces for improved cancer-targeted nanotherapy. *J. Control. Release* **2022**, *348*, 127–147. https://doi.org/10.1016/j.jconrel.2022.05.055.
17. Bros, M.; Nuhn, L.; Simon, J.; Moll, L.; Mailänder, V.; Landfester, K.; Grabbe, S. The protein corona as a confounding variable of nanoparticle-mediated targeted vaccine delivery. *Front. Immunol.* **2018**, *9*, 1760. https://doi.org/10.3389/fimmu.2018.01760.
18. Sánchez-Moreno, P.; Ortega-Vinuesa, J.L.; Martín-Rodríguez, A.; Boulaiz, H.; Marchal-Corrales, J.A.; Peula-García, J.M. Characterization of different functionalized lipidic nanocapsules as potential drug carriers. *Int. J. Mol. Sci.* **2012**, *13*, 2405–2424. https://doi.org/10.3390/ijms13022405.
19. Graván, P.; Aguilera-Garrido, A.; Marchal, J.A.; Navarro-Marchal, S.A.; Galisteo-González, F. Lipid-core nanoparticles: Classification, preparation methods, routes of administration and recent advances in cancer treatment. *Adv. Colloid Interface Sci.* **2023**, *314*, 102871. https://doi.org/10.1016/j.cis.2023.102871.
20. Braun, T.; Márk, L.; Ohmacht, R.; Sharma, U. Olive Oil as a Biocompatible Solvent for Pristine C 60. *Fuller. Nanotub. Carbon Nanostructures* **2007**, *15*, 311–314. https://doi.org/10.1080/15363830701423914.
21. Bharmoria, P.; Bisht, M.; Gomes, M.C.; Martins, M.; Neves, M.C.; Mano, J.F.; Bdikin, I.; Coutinho, J.A.P.; Ventura, S.P.M. Protein-olive oil-in-water nanoemulsions as encapsulation materials for curcumin acting as anticancer agent towards MDA-MB-231 cells. *Sci. Rep.* **2021**, *11*, 9099. https://doi.org/10.1038/s41598-021-88482-3.
22. Navarro-Marchal, S.A.; Griñán-Lisón, C.; Entrena, J.-M.; Ruiz-Alcalá, G.; Tristán-Manzano, M.; Martin, F.; Pérez-Victoria, I.; Peula-García, J.M.; Marchal, J.A. Anti-CD44-Conjugated Olive Oil Liquid Nanocapsules for Targeting Pancreatic Cancer Stem Cells. *Biomacromolecules* **2021**, *22*, 1374–1388. https://doi.org/10.1021/acs.biomac.0c01546.
23. Calvo, P.; Remuñán-López, C.; Vila-Jato, J.L.; Alonso, M.J. Novel hydrophilic chitosan-polyethylene oxide nanoparticles as protein carriers. *J. Appl. Polym. Sci.* **1997**, *63*, 125–132. https://doi.org/10.1002/(SICI)1097-4628(19970103)63:1<125::AID-APP13>3.0.CO;2-4.
24. Santander-Ortega, M.J.; Lozano-López, M.V.; Bastos-González, D.; Peula-García, J.M.; Ortega-Vinuesa, J.L. Novel core-shell lipid-chitosan and lipid-poloxamer nanocapsules: Stability by hydration forces. *Colloid Polym. Sci.* **2010**, *288*, 159–172. https://doi.org/10.1007/s00396-009-2132-y.
25. Sánchez-Moreno, P.; Buzón, P.; Boulaiz, H.; Peula-García, J.M.; Ortega-Vinuesa, J.L.; Luque, I.; Salvati, A.; Marchal, J.A. Balancing the effect of corona on therapeutic efficacy and macrophage uptake of lipid nanocapsules. *Biomaterials* **2015**, *61*, 266–278. https://doi.org/10.1016/j.biomaterials.2015.04.049.
26. Díaz-Torres, R.; López-Arellano, R.; Escobar-Chávez, J.J.; García-García, E.; Domínguez-Delgado, C.L.; Ramírez-Noguera, P. Effect of Size and Functionalization of Pharmaceutical Nanoparticles and Their Interaction with Biological Systems. In *Handbook of Nanoparticles*; Springer International Publishing: Cham, Switzerland, 2016; pp. 1041–1060.
27. Panyam, J.; Labhasetwar, V. Biodegradable nanoparticles for drug and gene delivery to cells and tissue. *Adv. Drug Deliv. Rev.* **2003**, *55*, 329–347. https://doi.org/10.1016/S0169-409X(02)00228-4.

28. Elias, D.R.; Poloukhtine, A.; Popik, V.; Tsourkas, A. Effect of ligand density, receptor density, and nanoparticle size on cell targeting. *Nanomed. Nanotechnol. Biol. Med.* **2013**, *9*, 194–201. https://doi.org/10.1016/j.nano.2012.05.015.
29. Saha, B.; Songe, P.; Evers, T.H.; Prins, M.W.J. The influence of covalent immobilization conditions on antibody accessibility on nanoparticles. *Analyst* **2017**, *142*, 4247–4256. https://doi.org/10.1039/C7AN01424D.
30. Torcello-Gómez, A.; Santander-Ortega, M.J.; Peula-García, J.M.; Maldonado-Valderrama, J.; Gálvez-Ruiz, M.J.; Ortega-Vinuesa, J.L.; Martín-Rodríguez, A. Adsorption of antibody onto Pluronic F68-covered nanoparticles: Link with surface properties. *Soft Matter* **2011**, *7*, 8450. https://doi.org/10.1039/c1sm05570d.
31. Martin, A.; Puig, J.; Galisteo, F.; Serra, J.; Hidalgo-Alvarez, R. On Some Aaspects of the Adsorption of Immunoglobulin-G Molecules on Polystyrene Microspheres. *J. Dispers. Sci. Technol.* **1992**, *13*, 399–416. https://doi.org/10.1080/01932699208943324.
32. Peula, J.M.; Hidalgo-Alvarez, R.; De Las Nieves, F.J. Coadsorption of IgG and BSA onto sulfonated polystyrene latex: I. Sequential and competitive coadsorption isotherms. *J. Biomater. Sci. Polym. Ed.* **1996**, *7*, 231–240. https://doi.org/10.1163/156856295X00274.
33. Peula-Garcia, J.M.; Hidaldo-Alvarez, R.; De las Nieves, F.J. Protein co-adsorption on different polystyrene latexes: Electrokinetic characterization and colloidal stability. *Colloid Polym. Sci.* **1997**, *275*, 198–202. https://doi.org/10.1007/s003960050072.
34. Hong, X.; Meng, Y.; Kalkanis, S.N. Serum proteins are extracted along with monolayer cells in plasticware and interfere with protein analysis. *J. Biol. Methods* **2016**, *3*, e51. https://doi.org/10.14440/jbm.2016.129.
35. Lee, D.Y.; Lee, S.Y.; Yun, S.H.; Jeong, J.W.; Kim, J.H.; Kim, H.W.; Choi, J.S.; Kim, G.-D.; Joo, S.T.; Choi, I.; et al. Review of the Current Research on Fetal Bovine Serum and the Development of Cultured Meat. *Food Sci. Anim. Resour.* **2022**, *42*, 775–799. https://doi.org/10.5851/kosfa.2022.e46.
36. Mathew, J.; Sankar, P.; Varacallo, M. *Physiology, Blood Plasma*; StatPearls: Treasure Island, FL, USA, 2023.
37. Mohammad-Beigi, H.; Hayashi, Y.; Zeuthen, C.M.; Eskandari, H.; Scavenius, C.; Juul-Madsen, K.; Vorup-Jensen, T.; Enghild, J.J.; Sutherland, D.S. Mapping and identification of soft corona proteins at nanoparticles and their impact on cellular association. *Nat. Commun.* **2020**, *11*, 4535. https://doi.org/10.1038/s41467-020-18237-7.
38. Amici, A.; Caracciolo, G.; Digiacomo, L.; Gambini, V.; Marchini, C.; Tilio, M.; Capriotti, A.L.; Colapicchioni, V.; Matassa, R.; Familiari, G.; et al. In vivo protein corona patterns of lipid nanoparticles. *RSC Adv.* **2017**, *7*, 1137–1145. https://doi.org/10.1039/C6RA25493D.
39. Kari, O.K.; Ndika, J.; Parkkila, P.; Louna, A.; Lajunen, T.; Puustinen, A.; Viitala, T.; Alenius, H.; Urtti, A. In situ analysis of liposome hard and soft protein corona structure and composition in a single label-free workflow. *Nanoscale* **2020**, *12*, 1728–1741. https://doi.org/10.1039/C9NR08186K.
40. Ketrat, S.; Japrung, D.; Pongprayoon, P. Exploring how structural and dynamic properties of bovine and canine serum albumins differ from human serum albumin. *J. Mol. Graph. Model.* **2020**, *98*, 107601. https://doi.org/10.1016/j.jmgm.2020.107601.
41. Kopac, T. Protein corona, understanding the nanoparticle–protein interactions and future perspectives: A critical review. *Int. J. Biol. Macromol.* **2021**, *169*, 290–301. https://doi.org/10.1016/j.ijbiomac.2020.12.108.
42. Salvati, A.; Pitek, A.S.; Monopoli, M.P.; Prapainop, K.; Bombelli, F.B.; Hristov, D.R.; Kelly, P.M.; Åberg, C.; Mahon, E.; Dawson, K.A. Transferrin-functionalized nanoparticles lose their targeting capabilities when a biomolecule corona adsorbs on the surface. *Nat. Nanotechnol.* **2013**, *8*, 137–143. https://doi.org/10.1038/nnano.2012.237.
43. Bilardo, R.; Traldi, F.; Vdovchenko, A.; Resmini, M. Influence of surface chemistry and morphology of nanoparticles on protein corona formation. *Wiley Interdiscip. Rev. Nanomed. Nanobiotechnology* **2022**, *14*, e1788. https://doi.org/10.1002/wnan.1788.
44. Salvati, A.; Dawson, K.A. What does the cell see? *Nat. Nanotechnol.* **2009**, *4*, 546–547. https://doi.org/10.1038/nnano.2009.248.
45. Acuña, A.U.; González-Rodríguez, J.; Lillo, M.P.; Naqvi, K.R. Protein structure probed by polarization spectroscopy. *Biophys. Chem.* **1987**, *26*, 63–70. https://doi.org/10.1016/0301-4622(87)80008-X.
46. Faizullin, D.; Valiullina, Y.; Salnikov, V.; Zuev, Y. Direct interaction of fibrinogen with lipid microparticles modulates clotting kinetics and clot structure. *Nanomed. Nanotechnol. Biol. Med.* **2020**, *23*, 102098. https://doi.org/10.1016/j.nano.2019.102098.
47. Peula, J.M.; de las Nieves, F.J. Adsorption of monomeric bovine serum albumin on sulfonated polystyrene model colloids 3. Colloidal stability of latex—Protein complexes. *Colloids Surf. A Physicochem. Eng. Asp.* **1994**, *90*, 55–62. https://doi.org/10.1016/0927-7757(94)02889-3.
48. Wasilewska, M.; Adamczyk, Z.; Jachimska, B. Structure of Fibrinogen in Electrolyte Solutions Derived from Dynamic Light Scattering (DLS) and Viscosity Measurements. *Langmuir* **2009**, *25*, 3698–3704. https://doi.org/10.1021/la803662a.
49. Cukalevski, R.; Ferreira, S.A.; Dunning, C.J.; Berggård, T.; Cedervall, T. IgG and fibrinogen driven nanoparticle aggregation. *Nano Res.* **2015**, *8*, 2733–2743. https://doi.org/10.1007/s12274-015-0780-4.
50. Gref, R.; Lück, M.; Quellec, P.; Marchand, M.; Dellacherie, E.; Harnisch, S.; Blunk, T.; Müller, R. 'Stealth' corona-core nanoparticles surface modified by polyethylene glycol (PEG): Influences of the corona (PEG chain length and surface density) and of the core composition on phagocytic uptake and plasma protein adsorption. *Colloids Surf. B Biointerfaces* **2000**, *18*, 301–313. https://doi.org/10.1016/S0927-7765(99)00156-3.
51. Pozzi, D.; Colapicchioni, V.; Caracciolo, G.; Piovesana, S.; Capriotti, A.L.; Palchetti, S.; De Grossi, S.; Riccioli, A.; Amenitsch, H.; Laganà, A. Effect of polyethyleneglycol (PEG) chain length on the bio–nano-interactions between PEGylated lipid nanoparticles and biological fluids: From nanostructure to uptake in cancer cells. *Nanoscale* **2014**, *6*, 2782. https://doi.org/10.1039/c3nr05559k.
52. Rejman, J.; Oberle, V.; Zuhorn, I.S.; Hoekstra, D. Size-dependent internalization of particles via the pathways of clathrin- and caveolae-mediated endocytosis. *Biochem. J.* **2004**, *377*, 159–169. https://doi.org/10.1042/BJ20031253.

53. Niaz, S.; Forbes, B.; Raimi-Abraham, B.T. Exploiting Endocytosis for Non-Spherical Nanoparticle Cellular Uptake. *Nanomanufacturing* **2022**, *2*, 1–16. https://doi.org/10.3390/nanomanufacturing2010001.
54. Ma, S.; Gu, C.; Xu, J.; He, J.; Li, S.; Zheng, H.; Pang, B.; Wen, Y.; Fang, Q.; Liu, W.; et al. Strategy for Avoiding Protein Corona Inhibition of Targeted Drug Delivery by Linking Recombinant Affibody Scaffold to Magnetosomes. *Int. J. Nanomed.* **2022**, *17*, 665–680. https://doi.org/10.2147/IJN.S338349.
55. Kesharwani, P.; Banerjee, S.; Padhye, S.; Sarkar, F.H.; Iyer, A.K. Hyaluronic Acid Engineered Nanomicelles Loaded with 3,4-Difluorobenzylidene Curcumin for Targeted Killing of CD44+ Stem-Like Pancreatic Cancer Cells. *Biomacromolecules* **2015**, *16*, 3042–3053. https://doi.org/10.1021/ACS.BIOMAC.5B00941.
56. Qian, C.; Wang, Y.; Chen, Y.; Zeng, L.; Zhang, Q.; Shuai, X.; Huang, K. Suppression of pancreatic tumor growth by targeted arsenic delivery with anti-CD44v6 single chain antibody conjugated nanoparticles. *Biomaterials* **2013**, *34*, 6175–6184. https://doi.org/10.1016/J.BIOMATERIALS.2013.04.056.
57. Trabulo, S.; Aires, A.; Aicher, A.; Heeschen, C.; Cortajarena, A.L. Multifunctionalized iron oxide nanoparticles for selective targeting of pancreatic cancer cells. *Biochim. Biophys. Acta Gen. Subj.* **2017**, *1861*, 1597–1605. https://doi.org/10.1016/J.BBAGEN.2017.01.035.
58. Waks, A.G.; Winer, E.P. Breast Cancer Treatment: A Review. *JAMA* **2019**, *321*, 288–300. https://doi.org/10.1001/JAMA.2018.19323.
59. Su, G.; Jiang, H.; Xu, B.; Yu, Y.; Chen, X. Effects of Protein Corona on Active and Passive Targeting of Cyclic RGD Peptide-Functionalized PEGylation Nanoparticles. *Mol. Pharm.* **2018**, *15*, 5019–5030. https://doi.org/10.1021/acs.molpharmaceut.8b00612.
60. Wang, S.; Zhang, J.; Zhou, H.; Lu, Y.C.; Jin, X.; Luo, L.; You, J. The role of protein corona on nanodrugs for organ-targeting and its prospects of application. *J. Control Release* **2023**, *360*, 15–43. https://doi.org/10.1016/J.JCONREL.2023.06.014.
61. Xiao, W.; Gao, H. The impact of protein corona on the behavior and targeting capability of nanoparticle-based delivery system. *Int. J. Pharm.* **2018**, *552*, 328–339. https://doi.org/10.1016/J.IJPHARM.2018.10.011.
62. Bashiri, G.; Padilla, M.S.; Swingle, K.L.; Shepherd, S.J.; Mitchell, M.J.; Wang, K. Nanoparticle protein corona: From structure and function to therapeutic targeting. *Lab Chip* **2023**, *23*, 1432–1466. https://doi.org/10.1039/d2lc00799a.
63. Miranda-Hernández, M.P.; López-Morales, C.A.; Piña-Lara, N.; Perdomo-Abúndez, F.C.; Pérez, N.O.; Revilla-Beltri, J.; Molina-Pérez, A.; Estrada-Marín, L.; Flores-Ortiz, L.F.; Ruiz-Argüelles, A.; et al. Pharmacokinetic Comparability of a Biosimilar Trastuzumab Anticipated from Its Physicochemical and Biological Characterization. *Biomed Res. Int.* **2015**, *2015*, 874916. https://doi.org/10.1155/2015/874916.
64. Laemmli, U.K. Cleavage of Structural Proteins during the Assembly of the Head of Bacteriophage T4. *Nature* **1970**, *227*, 680–685. https://doi.org/10.1038/227680a0.
65. Brunelle, J.L.; Green, R. One-Dimensional SDS-Polyacrylamide Gel Electrophoresis (1D SDS-PAGE); In *Methods in Enzymology*; Academic Press: Cambridge, MA, USA, 2014; pp. 151–159.
66. Delgado-Calvo-Flores, J.M.; Peula-García, J.M.; Martínez-García, R.; Callejas-Fernández, J. Experimental Evidence of Hydration Forces between Polymer Colloids Obtained by Photon Correlation Spectroscopy Measurements. *J. Colloid Interface Sci.* **1997**, *189*, 58–65. https://doi.org/10.1006/jcis.1997.4815.